%% file: conference_101719.tex
\documentclass[conference]{IEEEtran}
\IEEEoverridecommandlockouts
\usepackage{cite}
\usepackage{amsmath,amssymb,amsfonts}
\usepackage{algorithmic}
\usepackage{graphicx}
\usepackage{textcomp}
\usepackage{xcolor}
\def\BibTeX{{\rm B\kern-.05em{\sc i\kern-.025em b}\kern-.08em
    T\kern-.1667em\lower.7ex\hbox{E}\kern-.125emX}}

\input{header}

\begin{document}

\title{Achieving increased Phasor POD performance\\by introducing a Control-Input Model}

\author{
\IEEEauthorblockN{Hallvar Haugdal, Kjetil Uhlen}
\IEEEauthorblockA{Department of Electric Power Engineering\\
Norwegian University of Science and Technology\\
Trondheim, Norway\\
\{hallvar.haugdal, kjetil.uhlen\}@ntnu.no}
\and
\IEEEauthorblockN{Hj\"{o}rtur J\'{o}hannsson}
\IEEEauthorblockA{Department of Electrical  Engineering \\
Technical University of Denmark\\
Kgs. Lyngby, Denmark\\
hjjo@elektro.dtu.dk}
}

\maketitle

\begin{abstract}
\input{00-abstract}
\end{abstract}

\begin{IEEEkeywords}
Power Oscillation Damping, FACTS, Phasor Estimation, Kalman Filtering
\end{IEEEkeywords}

\input{main}

\bibliographystyle{IEEEtran}
\bibliography{IEEEabrv, references}

\end{document}

%% file: header.tex
\usepackage{upgreek}
\usepackage{svg}
\usepackage{gensymb}
\usepackage{amsmath, amssymb, bm}
\usepackage{mathtools}
\usepackage{xprintlen}
\usepackage{comment}

\usepackage{array,tabularx}

\newenvironment{conditions*}
  {\par\vspace{\abovedisplayskip}\noindent
   \tabularx{\columnwidth}{>{$}l<{$} @{}>{${}}c<{{}$}@{} >{\raggedright\arraybackslash}X}}
  {\endtabularx\par\vspace{\belowdisplayskip}}
  
\newcommand\figwidth{0.875}
\newcommand\figcaptioncut{0.4cm}

%% file: 00-abstract.tex
In this paper, an enhancement to the well known Phasor Power Oscillation Damper is proposed, aiming to increase its performance. Fundamental to the functioning of this controller is the estimation of a phasor representing oscillatory behaviour at a particular frequency in a measured signal. The phasor is transformed to time domain and applied as a setpoint signal to a controllable device. The contribution in this paper specifically targets the estimation algorithm of the controller: It is found that increased estimation accuracy and thereby enhanced damping performance can be achieved by introducing a  prediction-correction scheme for the estimator, in the form of a Kalman Filter. The prediction of the phasor at the next step is performed based on the control signal that is applied at the current step. This enables more precise damping of the targeted mode.

The presented results, which are obtained from simulations on a Single-Machine Infinite Bus system and the IEEE 39-Bus system, indicate that the proposed enhancement improves the performance of this type of controller.

%% file: main.tex
\input{01-introduction}
\input{02-theory}

\input{03-results}
\input{04-discussion}
\input{05-conclusion}

%% file: 01-introduction.tex
\section{Introduction}

Electromechanical oscillations, although extensively investigated in the literature, is still a recurring problem in power grids around the globe 
\cite{Kosterev1999, ENTSO-E-2011, ENTSO-E-2017, ENTSO-E-2018}
. Trends like increasing penetration of renewable energy sources, increased electricity demand and increasing import/export of energy between countries cause larger and less predictable fluctuations in both generation and demand. This makes it more difficult to tune and design stabilizing controllers for power oscillation damping.

Installing Power System Stabilizers on most large synchronous machines is a very cost-effective way of mitigating low damped power oscillations \cite{Machowski2008}.
However, in some cases it has been beneficial to utilize separate controllable devices for efficient power oscillation damping purposes, such as Flexible AC Transmission Systems (FACTS). 

FACTS devices can be installed to perform specific functions, like load flow control, enhancing the usable transfer capacity, or mitigation of power oscillations \cite{Hingorani1999}. 
Installing new FACTS devices for the purpose of power oscillation damping is a costly investment.
The value of the investment depends on the performance of the algorithms responsible for generating the damping control signal, so it is of utmost importance that the algorithms are effective as possible. This motivates the research work presented in this paper, where an enhancement to the well known Phasor Power Oscillation Damper (P-POD) is proposed.

The P-POD was introduced in \cite{Angquist2001}, where it is used for mitigating inter-area oscillations by modulating a Thyristor Controlled Series Capacitor (TCSC). The P-POD functions by estimating a phasor representing the oscillations in a measured signal. The estimated phasor is phase shifted and used to generate a damping control signal, which modulates some controllable device. In \cite{Angquist2001}, Low Pass Filters (LPF) are used for the estimation. Numerous variants and enhancements of the P-POD have been proposed since its introduction: In \cite{Chaudhuri2010a}, a scheme for latency compensation and adaptive phase compensation is proposed, and Recursive Least Squares (RLS) is used for the estimation. Latency compensation is also the focus in \cite{Yu2018}, where the P-POD is used to control a Doubly Fed Induction Machine. In \cite{Beza2016}, a RLS estimator with variable forgetting factor is introduced, aiming to increase the phasor estimation accuracy during transient conditions.

In all the mentioned research works, the only source of information for the phasor estimator is the chosen output measurement, typically power or frequency. One additional signal that is definitely always available, which we argue should be taken into account in the estimation, is the control signal applied by the controller. We show that this can be achieved by introducing a prediction-correction scheme in the form of a Kalman Filter: In the prediction step, the amplitude and phase of the oscillations at the next time step is predicted based on the control signal applied by the controller. In the correction step, the measured signal is used to correct the predicted estimate. The expectancy is that this will facilitate damping of oscillations in a more controlled manner.

The proposed estimation scheme is developed by combining a state space representation of the linearized power system model with equations describing the P-POD estimator found in the literature. Not surprisingly, it is found that the prediction step can be performed if we assume that the transfer function residue of the targeted low damped mode is known (i.e. a single complex number), where the transfer function is from the applied control signal to the output measurement. Obtaining the required residue is not straight-forward in a large scale system, but an approximate estimate could be obtained from model based modal analysis or from a measurement-based technique.

The motivation for the presented research is twofold: First, for systems where a sufficiently accurate residue estimate can be obtained, the performance of the P-POD is expected to increase. Second, the proposed scheme opens up for further work towards a self-correcting estimator which continually adjusts the residue estimate in order to improve predictions, allowing the P-POD to be developed into a self-tuning/adaptive controller (this possibility being explored in ongoing work).

The derivation of the enhanced P-POD is presented in Section \ref{sec:STM}. In Section \ref{sec:Results}, results are presented, where the focus is on comparing the P-POD with and without the proposed enhancement. Further, the robustness of the enhanced controller against residue parameter error is investigated. Finally, discussion and conclusions are given in Sections \ref{sec:Discussion} and \ref{sec:Conclusion}, respectively.



%% file: 02-theory.tex
\section{Background}
    \label{sec:Background}
	Fundamental to the operating principle of the conventional P-POD is the separation of the measured signal $S(t)$ into an average component $\bar{S}$ and an oscillatory component, where the oscillatory component is represented by the phasor $\vec{S}$ \cite{Angquist2001}: 
	    	\begin{equation}
	    		S(t) = \bar{S} + \mathrm{Re}\{ \vec{S} e^{j\omega t} \}
			\label{eq:MeasurementEquation}
	    	\end{equation}
	Here, $\omega$ is the assumed frequency of the targeted low damped mode. The separation is achieved using LPFs \cite{Angquist2001}, or by an estimation algorithm like RLS \cite{Chaudhuri2010a}, \cite{Yu2018}, \cite{Beza2016}, or a  Kalman filter \cite{Chaudhuri2011}. Further, the control signal is generated by applying a suitable phase shift $\beta$ and a gain $K$ to the estimated phasor:
		\begin{equation}
			u(t) = \mathrm{Re} \left\{ K e^{j\beta} \left( \vec{S} e^{j\omega t} \right)\right\}
			\label{eq:ControlEquation}
		\end{equation}
	Finally, the control signal is applied by modulating the control setpoint (reference) of a controllable unit. In \cite{Angquist2001}, the measured signal is the power flow in a line, and the control signal modulates the reactance reference of a TCSC installed on the line. In this case, the compensation angle $\beta$ is set to $90\degree$.
	
	In general, assuming that a sufficiently accurate model of the system is available, the ideal phase compensation for any given measurement and control can be calculated from modal analysis. Considering a single input-single output system, the state space representation can be written as follows \cite{Kundur1994}:
	\begin{equation}
		\Delta\dot{\mathbf{x}} = \mathbf{A}\Delta\mathbf{x} + \mathbf{b}\Delta u
		\label{eq:StateSpace}
	\end{equation}
	\begin{equation}
		\Delta{y} = \mathbf{c}\Delta\mathbf{x}
		\label{eq:StateSpaceMeasurement}
	\end{equation}
Applying the modal transformation, we get the decoupled system,
	\begin{equation}
		\dot{\mathbf{z}} = \mathbf{\Lambda}\mathbf{z} + \mathbf{\Psi}\mathbf{b}\Delta{u}
		\label{eq:modal_analysis:decoupled_system}
	\end{equation}
where $\mathbf{\Lambda}$ is a diagonal matrix containing the eigenvalues of the system, i.e. $\mathbf{\Lambda} = \operatorname{diag}(\lambda_1, \lambda_2 \dots \lambda_n)$. Electromechanical oscillations are associated with complex eigenvalues $\lambda_i = \alpha_i \pm  j\omega_i$, from which the damping and frequency of oscillations can be determined. A given eigenvalue $\lambda_i$ has associated right and left eigenvectors $\boldsymbol{\upphi}_i$ and $\boldsymbol{\uppsi}_i$, respectively. From the eigenvectors, the transfer function residue associated with the mode can be calculated (as defined in e.g. \cite{Kundur1994}, \cite{Rogers2000}):
	\begin{equation}
		r_i=\mathbf{c}\boldsymbol{\upphi}_i \mathbf{\boldsymbol{\uppsi}}_i\mathbf{b}
    \end{equation}
Finally, the ideal phase compensation for the P-POD can be determined from the angle of the residue \cite{Yang1994}:
	\begin{equation}
	    \beta = 180^{\circ} - \arg \{r\}
	    \label{eq:PhaseCompensation}
	\end{equation}
It can be shown that a marginal feedback between the output and input with this phase shift moves the eigenvalue further into the left half-plane, thus increasing the damping.

\section{Enhancing the Estimation\\Algorithm of the Phasor POD }

\label{sec:STM}

\begin{figure}
    \centering
    \includegraphics[width=\figwidth\columnwidth]{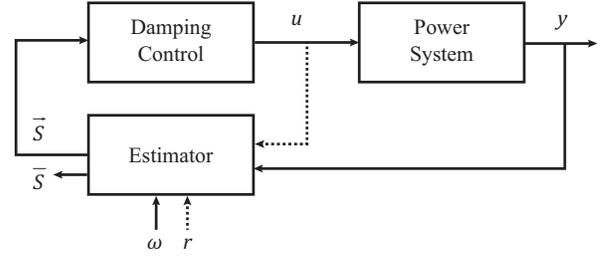}
    \caption{The figure shows a simplified block diagram of the P-POD. From the measured signal $y$, the "Estimator" block produces the estimate of the phasor $\vec{S}$. The "Damping Control" block rotates the phasor according to the desired phase compensation, and generates a time-domain control signal $u$. The dashed arrow indicates the additional information taken into account in the phasor estimation when introducing the proposed enhancement.}
    \label{fig:Flowchart}
    \vspace{-\figcaptioncut}
\end{figure}

In previous research on variants of the P-POD, the only information taken into account in the estimation of the phasor is the measured signal (based on \eqref{eq:MeasurementEquation}). The fundamental idea of the proposed enhancement is to also take the control signal applied by the P-POD into account in the phasor estimation. Specifically, the amplitude and angle of the phasor at the next time step is predicted based on the applied control signal at the current step. A prediction-correction estimator in the form of a Kalman Filter (KF) is suitable for this purpose. Fig. \ref{fig:Flowchart} shows a conceptual block diagram of the P-POD, where the proposed enhancement corresponds to adding the dashed arrow (i.e. feeding the control signal $u$ into the estimator block).

\subsection{Phasor Estimation: Prediction-Correction}

	A standard Kalman Filter \cite{BrownRobertGrover2012Itrs} can be given as follows,
    \begin{equation}
        \mathbf{X}_{k+1} =         \mathbf{F}_k \mathbf{X}_{k} + \mathbf{G}_k \mathbf{u}_k + \mathbf{w}_k
        \label{eq:KalmanFilterSTM}
    \end{equation}
   \begin{equation}
        \mathbf{Y}_k =  \mathbf{H}_k \mathbf{X}_k + \mathbf{v}_k
    \end{equation}
    where $\mathbf{X}_k$ is the Kalman Filter states, $\mathbf{F}_k$ is the State Transition Matrix, $\mathbf{G}_k$ is the Control-Input Model, $\mathbf{Y}_k$ is the measurement,
    $\mathbf{H}_k$ is the Observation Model,
    $\mathbf{w}_k$ is the process noise with covariance matrix $\mathbf{Q}$, and 
    $\mathbf{v}_k$ the measurement noise with covariance matrix $\mathbf{R}$.




 The Observation Model $\mathbf{H}_k$ relates the measurement to the states of the filter. The measurement is given by \eqref{eq:MeasurementEquation}, which can be written as follows:
	    	\begin{equation}
		\begin{split}
	    		S_k &=  \bar{S}_k + \mathrm{Re}\{ \vec{S}_k e^{j\omega t} \}\\
			&= \bar{S}_k +D_k \cos{\omega t} - Q_k \sin{\omega t}\\	
	    		&=
	    		\underbrace{
	    		\begin{bmatrix}
	    		1 & \cos{\omega t} & -\sin{\omega t}
	    		\end{bmatrix}
	    		}_{\mathbf{H}_k}
	    		\begin{bmatrix}
	    		\bar{S}_k \\
	    		D_k \\
	    		Q_k 
	    		\end{bmatrix}
		\end{split}
		\label{eq:P-POD-Obs}
	    	\end{equation}
This determines the observation model $\mathbf{H}_k$ for the Kalman filter, as indicated, 
	where the states are given by
	$
	    \mathbf{X}_k = 
	        \begin{bmatrix}
	    		\bar{S}_k, &
	    		D_k, &
	    		Q_k, 
	    	\end{bmatrix}^\intercal
	$, and the measurement by
	$
	    \mathbf{Y}_k = 
	    		S_k
	$. 

To determine the State Transition Matrix $\mathbf{F}_k$ and the Control-Input Matrix $\mathbf{G}_k$, we start by investigating how the phasor relates to the State Space representation of the linearized power system model given by \eqref{eq:StateSpace}, \eqref{eq:StateSpaceMeasurement} and \eqref{eq:modal_analysis:decoupled_system}.
Assuming that mode $m$ is the dominant oscillatory mode, the phasor we are attempting to estimate can be written as
\begin{equation}
	 \vec{S} e^{j\omega_m t} = Cz_m
	\label{eq:PhasorDefPre}
\end{equation}
where $C$ is a complex number determining the amplitude and phase of the oscillation in the measurement, relative to the modal response $z_m$.  To determine $C$ we relate the measured signal $S$ in \eqref{eq:MeasurementEquation} to the linearized system in \eqref{eq:StateSpaceMeasurement}:
\begin{equation}
\begin{split}
	S(t) &= y(t) = y_0 + \Delta{y}\\
	\bar{S} + \mathrm{Re}\{ \vec{S} e^{j\omega t} \} &= y_0 + \mathbf{c} \Delta{\mathbf{x}}\\
	 &= y_0 + \mathbf{c}\cdot \mathbf{\Phi}\mathbf{z}\\
	 &= y_0 + \mathbf{c}\cdot\sum_{	j\in \mathbb{A}} \boldsymbol{\upphi}_j z_j \\ 
\end{split}
\end{equation}

Here, we have used that $y(t) = y_0 + \Delta y$, and $\Delta\mathbf{x} = \mathbf{\Phi}\mathbf{z}$, where $\mathbf{\Phi}$ is the matrix of right eigenvectors.
$\boldsymbol{\upphi}_j$ is the right eigenvector (of dimension $n\times 1$, where $n$ is the order of the system) corresponding to mode $j$
. $\mathbb{A}$ is the set of indices of all modes (i.e. $\mathbb{A} = \{1 \dots n\}$). Introducing the sets $\mathbb{M} = \{m, \bar{m}\}$ (where $\bar{m}$ is the index of the complex conjugate of mode $m$) and $\mathbb{A}' = \mathbb{A} - \mathbb{M}$ allows us to separate out the terms associated with the targeted oscillations from the sum:
\begin{equation}
\bar{S} + \mathrm{Re}\{ \vec{S} e^{j\omega t} \} =
									 y_0 + \mathbf{c} \cdot\bigg[ \bigg( \sum_{		
							j\in\mathbb{A}'
									} 	 \boldsymbol{\upphi}_j z_j \bigg) +  \boldsymbol{\upphi}_m z_m +  \boldsymbol{\upphi}_{\bar{m}} z_{\bar{m}} \bigg]
\end{equation}
We would like to capture oscillations associated with mode $m$ with the phasor, so we assemble terms as follows:
\begin{equation}
	\mathrm{Re}\{ \vec{S} e^{j\omega t} \} 	= 2 \mathbf{c} \cdot \mathrm{Re}\{\boldsymbol{\upphi}_m z_m\}
\label{eq:PhasorDefRe}
\end{equation}

\begin{equation}
\bar{S} = y_0 +  \mathbf{c}\cdot\sum_{	
j \in \mathbb{A}'
}
\boldsymbol{\upphi}_j z_j
\label{eq:AvgDef}
\end{equation}
As shown in the second equation, the remaining terms not associated with the targeted oscillations are collected in the average $\bar{S}$. 
Combining \eqref{eq:PhasorDefPre} and \eqref{eq:PhasorDefRe}, we see that
\begin{equation}
	\vec{S} e^{j\omega t} =2 \mathbf{c}  \boldsymbol{\upphi}_m z_m
	\label{eq:PhasorDef}
\end{equation}
%
%
%
%
We now have an expression relating the phasor $\vec{S}$ to the modal response $z_m$, which can be combined with the solution of the decoupled state space system. The decoupled, discretized system can be written as follows:
            \begin{equation}
                \mathbf{z}_{k+1} =e^{\mathbf{\Lambda} \Delta t} \mathbf{z}_{k} +\mathbf{\Lambda}^{-1} \left(e^{\mathbf{\Lambda} \Delta t} - \mathbf{I}\right) \left(\mathbf{ \Psi b} \right) \Delta {u}_k
                \label{eq:discrete_state_space}
            \end{equation}        
Here, the subscript $k$ denotes the time step, and $\mathbf{\Psi}$ is the matrix of left eigenvectors. In the decoupled system, we can consider mode $m$ independently:
	\begin{equation}
          	z_{m, k+1} = e^{\lambda_m \Delta t} z_{m, k} + \frac{ \boldsymbol{\uppsi}_m \mathbf{b} }{\lambda_m} \left(e^{\lambda_m \Delta t} - 1\right) \Delta u_k
		\label{eq:ModeODESoln}
	\end{equation}
	Here, $\boldsymbol{\uppsi}_m$ is the the left eigenvector (of dimension $1\times n$) corresponding to the low damped mode. Since the damping is low it can be assumed that the real part of the eigenvalue is zero, i.e. $\lambda_m = j\omega_m$.
In the following, since we are focusing on a single mode only, we skip the index $m$ (i.e $z=z_m, \boldsymbol{\upphi} = \boldsymbol{\upphi}_m, \boldsymbol{\uppsi} = \boldsymbol{\uppsi}_m, \omega = \omega_m$). We introduce discrete notation also in \eqref{eq:PhasorDef}:
        \begin{equation}
             \vec{S}(t)e^{j\omega t} = 2\mathbf{c}\boldsymbol{\upphi} z(t)
             \Leftrightarrow
             \vec{S}_k e^{j\omega t} = 2 \mathbf{c}\boldsymbol{\upphi} z_k
            \label{eq:ModePhasorDef}
        \end{equation}
	Here, and in the following, $t=t_k$. Combining \eqref{eq:ModeODESoln} and \eqref{eq:ModePhasorDef} gives:
        \begin{equation}
            \vec{S}_{k+1}e^{j\omega (t + \Delta t)} = e^{j\omega \Delta t} \vec{S}_{k}e^{j\omega t} + \frac{2 \mathbf{c}\boldsymbol{\upphi} \mathbf{\boldsymbol{\uppsi}}\mathbf{b}}{\lambda} \left(e^{\omega \Delta t} - 1\right) \Delta u_k
        \end{equation}
        This expression can be written on the form
        \begin{equation}
                \vec{S}_{k+1} = \vec{S}_{k} + r \left( g(t) - jh(t) \right)\Delta u_k
        \end{equation}
        where we have introduced the residue $
		r=\mathbf{c}\boldsymbol{\upphi} \mathbf{\boldsymbol{\uppsi}}\mathbf{b}
    	$
	and defined the functions
        \begin{equation}
			g(t)=\frac{2}{\omega} \Big[- \sin\left(\omega t\right) + \sin\left(\omega(t+\Delta t)\right) \Big]
			\label{eq:g}
		\end{equation}
		\begin{equation}
			h(t)=\frac{2}{\omega} \Big[\cos\left(\omega t\right) - \cos\left(\omega(t+\Delta t)\right)\Big]
			\label{eq:h}
		\end{equation}
    	Further, we define the real and imaginary components of the residue:
    	\begin{equation}
    	    r = U + jV
    	    \label{eq:Ctrl}
    	\end{equation}
	Finally, we arrive at the prediction equations for the real and imaginary components of the phasor:
    	\begin{equation}
    	    D_{k+1} = D_k +  \left(Ug(t) + Vh(t) \right)\Delta u_k
    	    \label{eq:PhasorDModel}
    	\end{equation}
    	\begin{equation}
    	    Q_{k+1} = Q_k +  \left(-Uh(t) + Vg(t) \right)\Delta u_k
    	    \label{eq:PhasorQModel}
    	\end{equation}
    	The two equations \eqref{eq:PhasorDModel} and \eqref{eq:PhasorQModel} are used to predict the states of the filter at the next time step. For the average $\bar{S}_k$ we have no model (which does not require us to know the complete system model), so the best prediction is that the average remains unchanged, i.e. $\bar{S}_{k+1} = \bar{S}_k$. The prediction model on the form given by \eqref{eq:KalmanFilterSTM} becomes as follows: 
        \begin{equation}
		\begin{split}
    	    \begin{bmatrix}
        	    \bar{S}_{k+1}\\
        	    D_{k+1}\\
        	    Q_{k+1}\\
    	    \end{bmatrix}
    	    =
    	    \underbrace{
    	    \begin{bmatrix}
        	    1 &                     &    \\
        	    & 1    &                     \\
        	    & &                     1    \\
    	    \end{bmatrix}
    	    }_{\mathbf{F}_k}
    	    \begin{bmatrix}
    	        \bar{S}_{k}\\
        	    D_{k}\\
        	    Q_{k}\\
    	    \end{bmatrix}
    	    +
    	    \underbrace{
            \begin{bmatrix}
        	    0\\
        	    Ug(t) + Vh(t)\\
        	    -Uh(t) + Vg(t)\\
    	    \end{bmatrix}
    	    }_{\mathbf{G}_k}
    	    u_k
		\end{split}
		\label{eq:P-POD-X}
    	\end{equation}


This determines the State Transition Model $\mathbf{F}_k$ and the Control-Input Model $\mathbf{G}_k$. $\mathbf{G}_k$ is a function of time, and needs to be recalculated for each time step. 

To summarize: The Kalman filter as the P-POD estimator is given by the Observation Model in \eqref{eq:P-POD-Obs} and the State Transition- and Control-Input Model in \eqref{eq:P-POD-X}. The required parameters for the filter are the frequency of the targeted mode $\omega$ (in rad/s) and the residue $r=U+jV$. The frequency of the mode is easy to obtain, while the residue can be obtained either from modal analysis or from a measurement-based technique.
		
\subsection{Kalman Filter-based estimator vs Low Pass Filter- or Recursive Least Squares-based estimator}
\label{sec:PracticalConsiderations}

In \cite{Angquist2001}, where the P-POD was first presented, the phasor estimation is carried out using LPFs. For estimating a 0.2 Hz phasor, a cut-off frequency of 0.06 Hz is used for the filters, and in general a cut-off frequency of 0.2 to 0.5 times the targeted frequency is advised. A higher cut-off permits faster fluctuations in the estimated average and phasor components, while a lower cut-off gives a slower response.

The principal difference between the proposed phasor estimator given by \eqref{eq:P-POD-X} and previous LPF- or RLS-based variants is the control-input model $\mathbf{G_k}$. If no control action is applied, or if the residue estimate is zero, the Kalman Filter-based estimator is expected to behave similarly as the LPF- or RLS-based estimator. However, this requires that the Kalman Filter covariances are tuned properly. A starting point for the tuning of the Kalman filter can be obtained by comparing the response of the Kalman filter-based estimator with the LPF-based estimator. To achieve similar performance, the comparison is carried out without applying damping control (thus eliminating the effect of the Control-Input Model $\mathbf{G}_k$). By trial and error it is found that very similar performance is achieved with the two estimators when using the following covariance matrices:
\begin{equation}
    \mathbf{R} = 1, \quad \mathbf{Q} = \left(2\pi f \cdot \Delta t  \cdot k_c\right)^2\cdot \mathbf{I}_{3}
    \label{eq:Covariances}
\end{equation}
Here, $\Delta t$ is the time step, $f$ is the mode frequency and $k_c$ is the ratio of the LPF cut-off frequency to the mode frequency (typically in the range 0.2 to 0.5, as mentioned). 
This similarity is present also when varying the targeted mode frequency $f$, the time step $\Delta t$ or the cut-off frequency to mode frequency ratio $k_c$. Multiplying both covariance matrices $\mathbf{R}$ and $\mathbf{Q}$ by the same factor does not affect the performanc.

This is a good starting point for tuning of the Kalman filter. Similarly as with the LPF-based P-POD estimator, as discussed above, the Kalman filter-based estimator can be tuned by specifying the time step, the frequency of the targeted mode and selecting a value for $k_c$ in the range 0.2 to 0.5. Again, higher values allow faster fluctuations, while lower values result in a slower response.




For the initial conditions for the Estimate Covariance Matrix, commonly denoted by $\mathbf{P}_0$, higher values causes quicker convergence, but could lead to erratic control actions immediately after starting the estimator. 
Here, we have initialized this matrix as follows:
\begin{equation}
    \mathbf{P}_0 = 10^4 \cdot 2 \pi f \cdot \Delta t\cdot  k_c  \cdot  \mathbf{I}_{3}  
\end{equation}
Experience indicates that this choice does not affect results to a great deal, except at the first few hundreds of milliseconds of the simulation.

%% file: 03-results.tex
\section{Results}
\label{sec:Results}

In this section, the focus is comparing the performance of the P-POD \textit{with} and \textit{without} the proposed enhancement. To emphasize the performance advantage of introducing the proposed enhancement, the P-POD in its simplest form is considered. Frequency correction or other enhancements found in the literature, like adaptive forgetting factor, adaptive phase compensation, latency compensation etc. are not considered. It should be mentioned, however, that most, if not all of these enhancements are compatible with the proposed enhancement.

In the following, the conventional LPF- or RLS-based P-POD, without a Control-Input Model, is considered as the reference case, and is referred to as P-POD-0. The enhanced P-POD, with the proposed Control-Input Model (CIM), is referred to as P-POD-CIM.


As mentioned in \cite{Angquist2001}, the main motivation for installing FACTS devices is the mitigation of severe oscillations following major disturbances. Therefore, controllers are tested on standing oscillations or low damped ringdowns, rather than e.g. using random load variations for exciting the system.

All simulations are carried out in Python using a simulation package described in \cite{DynPSSimPy}, which was developed specifically for research work towards enhancing the P-POD. Generators are represented by 6\textsuperscript{th} order generator model given in \cite{Machowski2008}, and all currents and voltages are represented by phasors. The TCSC is modelled as described in \cite{Pal2005}. Integration of differential equations is performed with a constant time step size of $5$ ms using the Modified Euler method with one correction iteration. The Kalman Filters are updated every $20$ ms.

\subsection{Single-Machine Infinite Bus}
The synchronous machine parameters are based on the case described in \cite[p.~752]{Kundur1994}, but leakage reactance, armature resistance and saturation are neglected. The machine is equipped with a simple excitation system (AVR model SEXS). The P-POD measures the generator speed deviation, and modulates the reactance reference of a TCSC installed on the line connecting the synchronous machine and the infinite bus. Following the example established in \cite{Chaudhuri2004}, the TCSC has a steady state compensation of 10\%, and minimum and maximum compensation limits of 1\% and 50\%, respectively. The simulated event is a short circuit with a clearing time of 50 ms applied on the terminals of the synchronous machine.

Through modal analysis on the linearized model, it is found that this system has an unstable electromechanical eigenvalue
with a frequency of $1.01$ Hz and damping of $-2.09\%$. The residue corresponding to the measurement, control actuator and mode is $r=0.036\angle158\degree$, from which the
phase compensation is determined (according to \eqref{eq:PhaseCompensation}), i.e. $\beta= 180\degree - 158\degree = 22\degree$. The residue also determines the coefficients $U$ and $V$ in \eqref{eq:P-POD-X}.


\begin{figure}
    \centering
    \includegraphics[width=\figwidth\columnwidth]{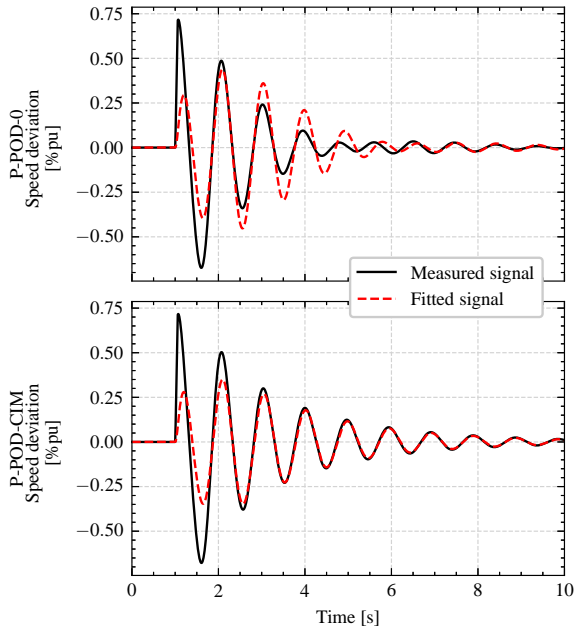}
    \caption{The figure shows the results from testing the P-POD-0 (upper plot) and the P-POD-CIM (lower plot) on the Single-Machine Infinite Bus-system. The measured signal is shown along the estimated oscillations, revealing that the accuracy is higher with the P-POD-CIM.}
    \label{fig:SMIB-STMCompEqGain}
    \vspace{-\figcaptioncut}
\end{figure} 

\subsubsection{Comparison - Equal gain}
Fig. \ref{fig:SMIB-STMCompEqGain} shows the results from two simulations performed on the above described system, where the P-POD-0 (upper) and the P-POD-CIM (lower) are used to generate the damping signal. The gain used in both cases is 15. The fitted signals are shown along the measured signals, and comparing the two plots shows that a more accurate fit is achieved with the P-POD-CIM. The two cases can be compared in terms of control cost $C$ and performance $P$, which can be defined as follows:

\begin{equation}
    C = \sqrt{\sum_{k=0}^{N-1} u_k^2}, \quad P = \left( \sqrt{\sum_{k=0}^{N-1} \Delta x_k^2} \right)^{-1}
    \label{eq:CostPerformance}
\end{equation}
where $N$ is the number of time steps, $u_k$ is the control action, $\Delta x_k$ is the generator speed deviation and $k$ is the time step.  Both the cost ($\approx22\%$) and the performance ($\approx7\%$) are higher with the P-POD-0 than with the P-POD-CIM. However, for a fair comparison, the gain should be adjusted such that the control cost is the same in the two cases.

\subsubsection{Comparison - Varying gain}

\begin{figure}
    \centering
    \includegraphics[width=\figwidth\columnwidth]{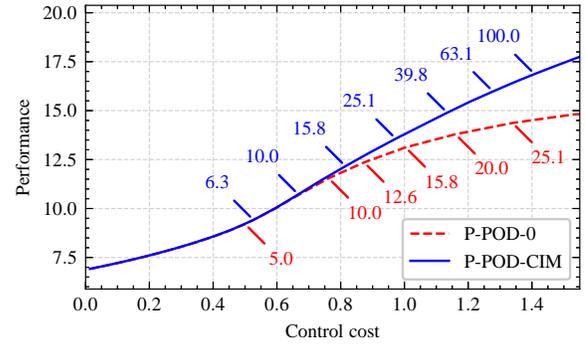}
    \caption{The performance and control cost for varying gains are shown for the P-POD-0 and the P-POD-CIM. The gain is indicated along the curves. The result indicates that the P-POD-CIM has a higher performance than the P-POD-0 for the same control cost.}
    \label{fig:SMIB-STMCompGain}
    \vspace{-\figcaptioncut}
\end{figure}

To compare the P-POD-0 and the P-POD-CIM on equal terms, the same two simulations are performed for a range of gains between 0 and 100. The cost and performance are calculated using \eqref{eq:CostPerformance}, and plotted in Fig. \ref{fig:SMIB-STMCompGain}. This result clearly shows that for the same control cost, the P-POD-CIM performs better than the P-POD-0.

\subsubsection{Comparison - Equal cost}

\begin{figure}
    \centering
     \includegraphics[width=\figwidth\columnwidth]{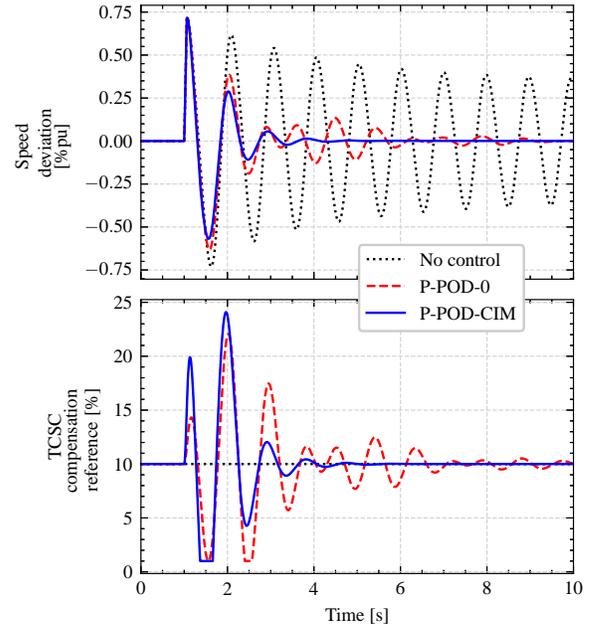}
    \caption{A comparison of three cases on the Single-Machine Infinite Bus system shown: No control, P-POD-0 and P-POD-CIM. The gains are adjusted so the control cost is about the same in the two cases with control (28 and 100 respectively). In both cases with control the oscillations are stabilized, but the case with P-POD-CIM is clearly preferable.}
    \label{fig:SMIB-STMCompEqCost}
    \vspace{-\figcaptioncut}
\end{figure}

From Fig. \ref{fig:SMIB-STMCompGain}, it is observed that the control cost is approximately the same when the P-POD-0 is applied with a gain slightly above 25, and the P-POD-CIM is applied with a gain of 100. Fig. \ref{fig:SMIB-STMCompEqCost} shows a comparison of the two controls under these circumstances (the actual gain used for the P-POD-0 is 28, resulting in a 1\% higher control cost for the P-POD-0 than for the P-POD-CIM). For reference, the unregulated system is also shown, which exhibits standing oscillations. The performance is 15\% better for the P-POD-CIM in this case. Also, looking at the curves, it is clear that the P-POD-CIM is preferable to the P-POD-0.

\subsection{Effect of Residue Parameter Error}

\begin{figure}
    \centering
    \includegraphics[width=\figwidth\columnwidth]{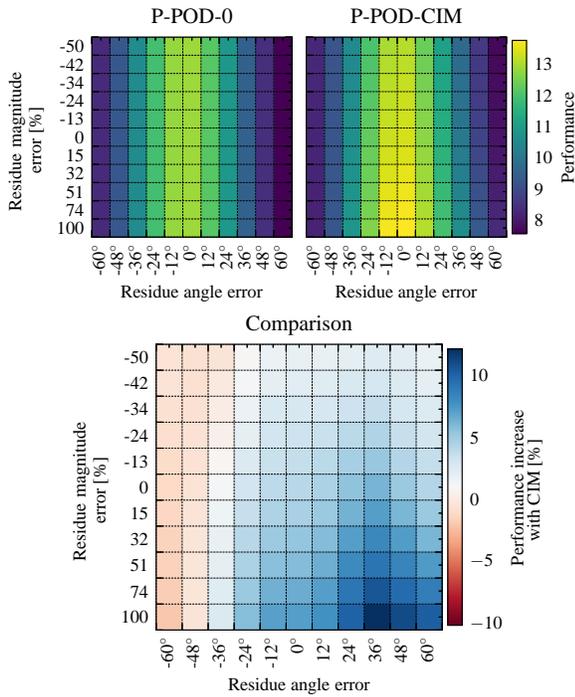}
    \caption{The figure shows a performance comparison of the P-POD-0 and the P-POD-CIM, operating with non-optimal phase compensation and inaccurate Control-Input Model. The performance of the P-POD-0 is shown to the upper left, and the P-POD-CIM to the upper right. The control cost required to achieve the given performance is equal in all cases. The percentage-wise performance advantage is shown in the lower plot.}
    \label{fig:ResidueError}
    \vspace{-\figcaptioncut}
\end{figure}

The above results indicates that the P-POD-CIM provides a more effective damping control signal than the P-POD-0, provided accurate knowledge of the required residue. However, an accurate residue value might not be straightforward to obtain in practice. Furthermore, the residue will likely change with changing operating conditions. An important next step in the analysis is therefore to investigate how an inaccurate residue affects the performance of the P-POD-CIM.

Assuming that some inaccurate residue estimate is provided, either from model based modal analysis or from a measurement-based method, the goal is to assess whether the P-POD-0 or the P-POD-CIM will perform better. Again, to make a comparison on equal terms, the performance of the controllers should be compared at the same cost. However, instead of looking at the full curve, as in Fig. \ref{fig:SMIB-STMCompGain}, one specific control cost is chosen where the performance is evaluated.

The same system and scenario is analysed for various test residues. The test residues are generated by scaling and/or rotating the exact residue. Scaling values are chosen between 0.5 and 2, and rotation angles between $-60^{\circ}$ and $+60^{\circ}$. This results in a 2D-grid of test residues.

For each test residue, a number of simulations with varying gains are simulated, both with the P-POD-0 and the P-POD-CIM (similarly as in Fig. \ref{fig:SMIB-STMCompGain}). For the P-POD-0, the phase compensation $\beta$ is determined from the residue, and for the P-POD-CIM, the residue determines both the phase compensation $\beta$ and the parameters $U$ and $V$. Further, the performance achieved at a specific control cost is found by interpolation (for example, considering Fig. \ref{fig:SMIB-STMCompEqGain}, which corresponds to zero residue error, the performance achieved at control cost 1.0 is about 13.7 for P-POD-CIM, and about 13.1 for P-POD-0). Finally, the percentage improvement of the P-POD-CIM relative to the P-POD-0 can be found by $(P_{CIM} - P_{0})/P_{CIM}\cdot 100\%$.

The result from the above procedure is presented in Fig. \ref{fig:ResidueError}. The plot to the upper left shows the performance achieved with the P-POD-0. Since the phase compensation for the P-POD-0 is chosen based on the residue angle, the performance varies with residue angle error. As expected, the performance is maximised when $\beta$ is chosen according to \eqref{eq:PhaseCompensation}. The residue magnitude does not affect the performance of the P-POD-0.

For the P-POD-CIM, shown in the upper right plot, the performance varies both with angle error and magnitude error. Interestingly, the performance increases when the magnitude of the test residue is larger than the exact residue.

The lower plot indicates that the performance achieved with the P-POD-CIM is higher than that achieved with the P-POD-0 for test residues within a deviation of $-24^{\circ}$ and $+60^{\circ}$ from the exact residue.




\subsection{Application to the IEEE 39-Bus System}

\begin{figure}
    \centering
    \includegraphics[width=\columnwidth]{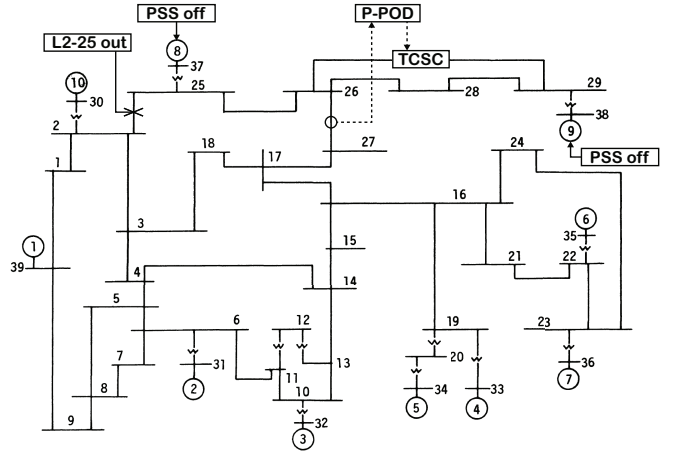}
    \caption{The IEEE 39 bus system is shown. The P-POD measures the power flow in the line between buses 26 and 27, and modualtes the reactance reference for the TCSC installed on the line between buses 26 and 29.}
    \label{fig:my_label}
    \vspace{-\figcaptioncut}
\end{figure}


\begin{figure}
    \centering
\includegraphics[width=\figwidth \columnwidth]{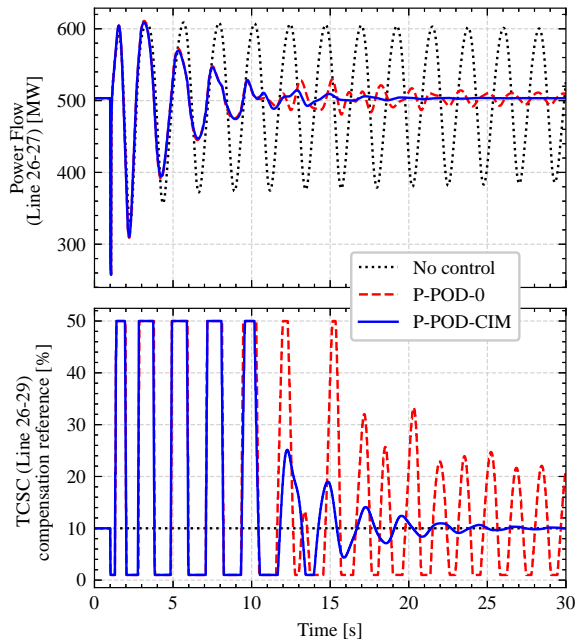}
    \caption{The performance of the P-POD-0 is compared with the P-POD-CIM on the IEEE 39 bus system. The performance of the P-POD-CIM is superior in this case.}
    \label{fig:IEEE39-STMComp}
    \vspace{-\figcaptioncut}
\end{figure}
A final comparison of the P-POD with and without Control-Input Model is made based on simulations on the IEEE 39 Bus System \cite{Pai1989}, which is a significantly larger and more complex system than the Single-Machine Infinite Bus system. This system consists of 39 buses and 10 generators. All generators except the large machine (generator 1) representing the rest of USA and Canada are equipped with AVR, turbine-governor and PSS controls. A TCSC is installed on the line between buses 26 and 29, which is modulated by the P-POD. The input measurement for the P-POD is the active power flow in the line between buses 26 and 27, and the gain is 20. An unstable operating condition is provoked by disconnecting the line between buses 2 and 25 and deactivating the stabilizers on generators 8 and 9. By modal analysis it is found that the system is unstable with a $0.44$ Hz mode with zero damping. 

The simulated event is a short-circuit on bus 2, with a clearing time of $50$ ms. The result in Fig. \ref{fig:IEEE39-STMComp} shows that only the P-POD-CIM performs satisfactorily. The P-POD-0 applies unwanted control action after the targeted oscillations have died out. It should be mentioned that for lower gains both controllers perform similarly, but this result indicates that a higher gain is permitted with the P-POD-CIM without producing unwanted effects.

%% file: 04-discussion.tex
\section{Discussion}
\label{sec:Discussion}
Three important conclusions can be drawn based on the presented results: First, the results indicate that the proposed enhancement to the P-POD increases the damping performance of the controller. This is as expected, given that more information is taken into account in the phasor estimation, and that more knowledge of the system is required (in the form of the transfer function residue). Second, with the proposed enhancement, higher gains are permitted before the controller produces unwanted effects. Third, the enhanced estimator is robust against residue parameter error. This is important, since acquiring an accurate estimate of the required transfer function residue might not be straightforward.

An interesting possibility that opens up with the introduction of the proposed prediction-correction estimator scheme, is to develop the P-POD into an adaptive P-POD:  
Multiple Kalman filter-based P-POD estimators could be running in parallel, each making predictions based on a specific, assigned residue. 
Estimators with a high prediction accuracy would indicate that the corresponding residue was close to the actual residue, and vice versa for estimators with a low prediction accuracy. The phase compensation could then be adjusted according to the most accurate estimators, resulting in an adaptive/self-tuning P-POD responding to changing operating conditions. This is the topic of further research.

Finally, it should be mentioned that there are no expected challenges with running the enhanced P-POD in real-time. This is also verified by testing the controller in a real-time simulation environment.

%% file: 05-conclusion.tex
\section{Conclusion}
\label{sec:Conclusion}
An enhancement to the Phasor Power Oscillation Damper has been presented. The results indicate that the performance of this type of controllers can be improved with the proposed solution, based on analysis both for a Single-Machine Infinite Bus system and for the IEEE 39-Bus System. It is also found that the controller is robust against residue parameter error. Finally, the presented theory lays the foundation for developing the P-POD into an adaptive controller for mitigating power oscillations under changing operating conditions.